\documentclass{article}
\usepackage{spconf,amsmath,graphicx}
\usepackage{bbm}
\usepackage{booktabs}
\usepackage[misc]{ifsym}
\usepackage{bbding}

\usepackage{enumitem}
\setlist{nosep, leftmargin=14pt}

\usepackage{mwe} 

\title{Precise Few-shot Fat-free Thigh Muscle Segmentation in T1-weighted MRI}

\name{
\begin{tabular}{@{}c@{}}
Sheng Chen$^{\star}$, 
Zihao Tang$^{\star\diamond}$, 
Dongnan Liu$^{\star}$, 
Ch{\'e} Fornusek$^{\circ}$,
Michael Barnett$^{\diamond\vee}$, \\
Chenyu Wang$^{\diamond\vee}$,
Mariano Cabezas$^{\diamond}$, 
Weidong Cai$^{\star}$
\end{tabular}}
\address{
$^{\star}$ School of Computer Science, University of Sydney, Australia \\
$^{\diamond}$ Brain and Mind Centre, University of Sydney, Australia\\
$^{\circ}$ Discipline of Exercise and Sport Science, Faculty of Medicine and Health, \\ University of Sydney, Australia\\
$^{\vee}$ Sydney Neuroimaging Analysis Centre, Australia}

\begin{document}
\maketitle

\begin{abstract}
Precise thigh muscle volumes are crucial to monitor the motor functionality of patients with diseases that may result in various degrees of thigh muscle loss. T1-weighted MRI is the default surrogate to obtain thigh muscle masks due to its contrast between muscle and fat signals. Deep learning approaches have recently been widely used to obtain these masks through segmentation. However, due to the insufficient amount of precise annotations, thigh muscle masks generated by deep learning approaches tend to misclassify intra-muscular fat (IMF) as muscle impacting the analysis of muscle volumetrics. As IMF is infiltrated inside the muscle, human annotations require expertise and time. Thus, precise muscle masks where IMF is excluded are limited in practice. To alleviate this, we propose a few-shot segmentation framework to generate thigh muscle masks excluding IMF. In our framework, we design a novel pseudo-label correction and evaluation scheme, together with a new noise robust loss for exploiting high certainty areas. The proposed framework only takes $1\%$ of the fine-annotated training dataset, and achieves comparable performance with fully supervised methods according to the experimental results.

\end{abstract}
\section{INTRODUCTION}
\label{sec:intro}
Precise thigh muscle volumes are critical for analyzing the motor functionality of patients with diseases that can cause different levels of muscle loss directly or indirectly due to sedentarism, such as amyotrophic lateral sclerosis~\cite{ALS} or multiple sclerosis~\cite{fornusek2014neuromuscular}. T1-weighted MRI is a commonly used surrogate for studying muscles as water is their primary mass component and T1-weighted images have good contrast between water and fat~\cite{lorenzo2019role}. Traditional thigh muscle segmentation approaches usually involve intensity-based algorithms and a relatively complex pipeline for pre- and post-processing the images~\cite{orgiu2016automatic}. Recently, with the advancements in deep learning strategies, convolutional neural networks (CNNs)~\cite{Unet, AttU, resU} have improved the accuracy of medical segmentation tasks and helped reduce human workload significantly~\cite{Unet, UDA}. However, when precise annotations are limited or noisy, the model performance decreases significantly~\cite{huang2021co}. In addition, the variability of the acquisition parameters for different scanners and protocols limits the generalization of learning-based methods~\cite{liu2020unsupervised}. Fully-supervised models may overfit a particular dataset and suffer from performance drops on other datasets. Fine-tuning to these other datasets is also constrained by the availability of the scans from these new acquisition setups. Regarding thigh muscle segmentation, distinguishing intra-muscular fat (IMF) from thigh muscle remains challenging. IMF is defined as the fat that appears between muscle tissues, and it usually presents a complex structure that is spread into discontinuous branches~\cite{IMF, LiteImf}. Furthermore, muscle regions tend to present intensity inhomogeneities that disrupt their intensity distribution and affect the segmentation results~\cite{orgiu2016automatic}. Thus, precise human annotation requires expertise and intensive labor, limiting the availability of precise thigh muscle masks where IMF is excluded.

To exploit the limited availability of precise annotations, few-shot segmentation methods commonly use a strategy based on class prototype information~\cite{wang2019panet}. Few-shot methods also address the performance drop problem between different domains by adapting to a new dataset with few scans. Consequently, semi-supervised methods have gained increasing attention in medical image segmentation to segment challenging areas and structures for specific clinical tasks~\cite{wu2022mutual}. For our purpose, pseudo-labels derived from precisely labeled data can enrich the datasets, improving the exclusion of IMF. Several works have shown that combining uncertainty estimation with semi-supervised tasks helps reduce noise in pseudo-labels~\cite{McNet, Uncertanity}. There are two popular approaches regarding uncertainty estimation to improve pseudo-labels. The first approach is to embed dropout layers~\cite{Uncertanity} and use Monte Carlo sampling to generate uncertainty maps to reduce the effect of poorly determined regions. The second approach is to construct multiple decoders with different up-sampling methods~\cite{wu2022mutual,zheng2021rectifying} and enforce a mutual consistency on the multiple outputs to constrain their uncertainty. 

In this paper, we propose a two-stage few-shot segmentation framework specifically designed for precise thigh muscle segmentation excluding IMF. The proposed framework includes a pseudo-label generation (PLG) stage and a pseudo-label denoising (PLD) stage. In the PLG stage, pseudo-labels are generated by MC-Net~\cite{McNet} to enrich the dataset. In the PLD stage, these pseudo-labels are denoised by a novel correction and evaluation strategy. Moreover, a noise-robust loss is proposed to exploit the corrected pseudo-labels. Evaluated by dice for IMF and thigh muscle, our results demonstrate that the proposed method using only $13$ ($1\%$ of the dataset) precisely labeled images outperforms other methods and can achieve comparable performance to the fully-supervised method. 

\section{METHOD}
\label{sec:METHOD}
\begin{figure}[htb]
  \centering
  \centerline{\includegraphics[width=8.5cm]{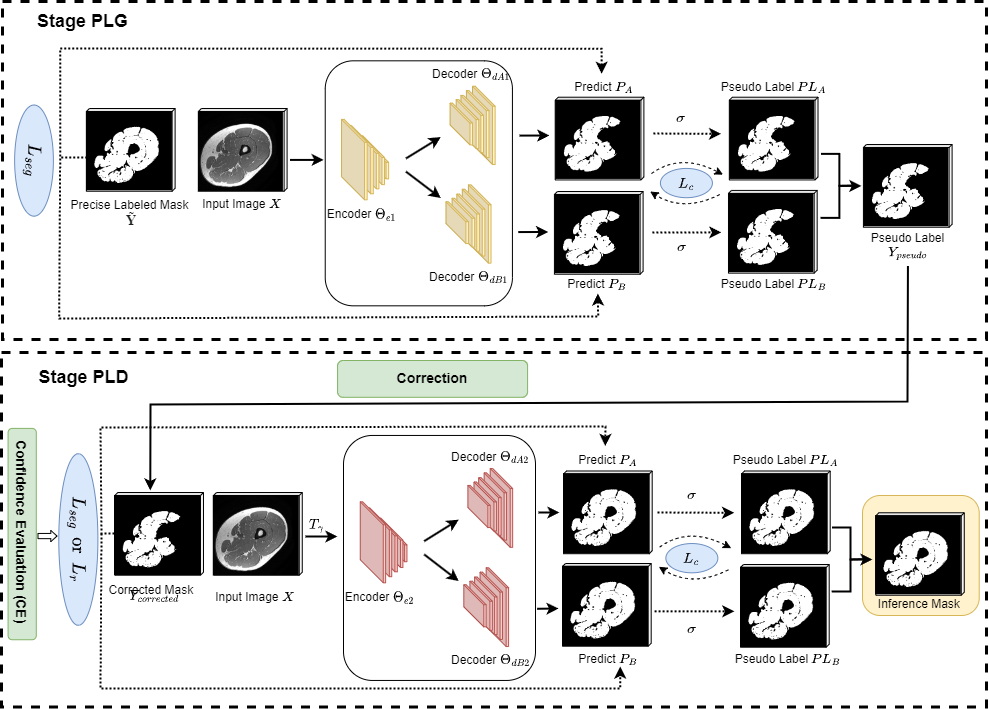}}

\caption{Overview of the proposed framework. Our framework follows a two-stage process of pseudo-label generation (PLG) and its posterior pseudo-label denoising (PLD).}
\label{fig:1}
\end{figure}

\noindent The overview of our two-stage framework is illustrated in Fig.~\ref{fig:1}. In the following sections, we focus on the two stages of our approach (PLG and PLD) and our novel loss function, define them formally and present the intuition behind them. Finally, we introduce our data augmentation scheme to further extend the training set.

\subsection{Pseudo-label Generation}
\label{sec:PLG}
Considering that thigh muscle annotations usually include the IMF, we chose MC-Net, a semi-supervised approach~\cite{McNet}, as our backbone to combine precise and noisy segmentations to enrich our training set. Specifically, two decoders ($\theta_{dA}$ and $\theta_{dB}$) are designed with different up-sampling strategies based on the features from a single encoder $\theta_e$. Two probability outputs ($P_A$ and $P_B$) are generated by independent decoders using the sigmoid function and then thresholded ($> 0.5$) into pseudo labels ($PL_A$ and $PL_B$). To ensure consistency and reduce uncertainty on the predictions, a cyclical scheme is introduced~\cite{wu2022mutual}.

Finally, the loss functions involved in this stage consist of a segmentation loss $L_{seg}$ and a self-supervised consistency loss $L_c$. $L_{seg}$ is only used for training samples with precisely annotated masks ($\Tilde{Y}$), where $L_c$ is used for all training data. The overall loss function can be formulated as:
\begin{gather}
    L_{seg} = Dice(P_A, \Tilde{Y}) + Dice(P_B,\Tilde{Y}), \\
    L_c = ||(P_A - PL_B)||_{2} + ||(P_B - PL_A)||_{2}.
\end{gather}
Once the model is trained, pseudo-labels ($Y_{pseudo}$) for all images ($X$) in the dataset can be generated by the mean prediction between the decoders as follows, where $\sigma$ is the sigmoid function followed by a thresholding operation:
\begin{equation}
\label{eq:5}
    Y_{pseudo} = \sigma (0.5 \times (\theta_{dA1}(\theta_{e1}(X)) + \theta_{dB1}(\theta_{e1}(X)))).
\end{equation}

\subsection{Pseudo-label Correction}
\label{sec:PLC}
Although MC-Net can reduce the uncertainty between the decoders, the generated pseudo-labels $Y_{pseudo}$ still contain errors, especially on challenging samples. To further distinguish thigh muscle from IMF and reduce the effect of the noisy pseudo-labels, we propose a new denoising strategy. 

Inspired by source-free domain adaptation approaches~\cite{UDA}, we argue that the feature within the same category should lie closer to their class prototypes, enforcing a high correlation for samples of the same class. In the case of segmentation, image regions with highly correlated voxels in feature space often have compact intensity ranges~\cite{shu2019lvc}. Thus, instead of estimating the distance of a pixel to different class prototypes~\cite{wang2019panet}, we directly exploit the class prototype information hidden in the intensity distribution of a set of given regions. As the intensity values of the thigh muscle are mostly distributed in the range of $(0.2, 0.6)$ after normalizing the input images ($X$) to the range $(0, 1)$, we extract coarse masks $Y_{coarse}$ that exclude most of the IMF tissue. A single coarse mask $Y_{coarse}$ is denoted as:
\begin{equation}
     Y_{coarse} = \begin{cases}
                    1, \text{pixel} \in (0.2, 0.6) \\
                    0, \text{otherwise}.
                 \end{cases}
\end{equation}

\begin{figure}[htb]
\centering
\begin{minipage}[b]{0.23\linewidth}
  \centering
  \centerline{\includegraphics[width=2.0cm]{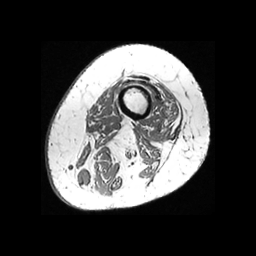}}
  \centerline{(a) Image $X$}\medskip
\end{minipage}
\hfill
\begin{minipage}[b]{0.23\linewidth}
  \centering
  \centerline{\includegraphics[width=2.0cm]{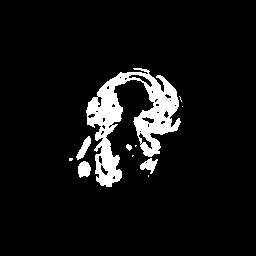}}
  \centerline{(b) $Y_{pseudo}$}\medskip
\end{minipage}
\hfill
\begin{minipage}[b]{0.23\linewidth}
  \centering
  \centerline{\includegraphics[width=2.0cm]{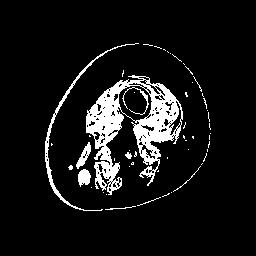}}
  \centerline{(c) $Y_{coarse}$}\medskip
\end{minipage}
\hfill
\begin{minipage}[b]{0.23\linewidth}
  \centering
  \centerline{\includegraphics[width=2.0cm]{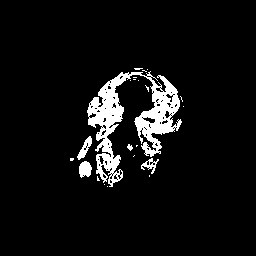}}
  \centerline{(d) $Y_{corrected}$}\medskip
\end{minipage}
\caption{Pseudo-label generation and correction examples. For an input image (a), we generate a pseudo-label mask (b) and intersect it with a coarse label mask (c) to produce the final corrected mask (d).}
\label{fig:2}
\end{figure}

Finally, we define the set of corrected pseudo-labels $Y_{corrected}$ as the intersection between the $Y_{coarse}$ and $Y_{pseudo}$ masks. Examples are shown in Fig.~\ref{fig:2}.

\subsection{Noise Robust Loss}
Even though the corrected pseudo-labels $Y_{corrected}$ can exclude most of the IMF, these masks present errors in the muscle's morphological structure. If the model is trained only with $Y_{corrected}$ masks, the original $L_{seg}$ will be affected by these errors and lead to incorrect masks. To address that issue, we propose a confidence evaluation (CE) strategy to ignore unreliable masks ($\Tilde{Y}_{corrected}$) when evaluating the $L_{seg}$ loss. Inspired by~\cite{wu2021novel}, we also introduce a new noise robust loss ($L_r$) that takes into account the $\Tilde{Y}_{corrected}$ masks.

Given an input image $X$ and its $Y_{pseudo}$ and $Y_{coarse}$ masks, we define its confidence score $S$ as the Dice similarity coefficient between the masks:
\begin{equation}
S =  2 \times \frac{(Y_{pseudo} \times Y_{coarse})}{Y_{pseudo} + Y_{coarse}}.
\end{equation}
In those cases where the confidence score is low, the resulting $Y_{corrected}$ mask will vary greatly from the original two masks due to their low intersection. This will either mean that the morphological structure of the muscle in $Y_{corrected}$ is unreliable or that the sample $X$ is a hard example. Consequently, if the confidence score of the $Y_{corrected}$ mask is lower than 0.8,  the sample is excluded from the $L_{seg}$ loss (now calculated on eligible $Y_{corrected}$ masks).

This new constrain, offers a trade-off between highly-confident structural segmentations and supervision information. To offset the effect of the loss examples and to fully exploit all the $Y_{corrected}$ pseudo-labels with missing muscle voxels, we design a loss that penalizes differences in the predictions ($P_A$ and $P_B$) under the $Y_{corrected}$ masks. Thus, we define our noise robust loss $L_r$ as:
\begin{multline}
    L_r = ||(P_A \times Y_{corrected}-Y_{corrected})||_{2} \\ + ||(P_B \times Y_{corrected}-Y_{corrected})||_{2}.
\end{multline}

\noindent Finally, the overall loss function for this second stage in our framework can be defined as: 
\begin{gather}
    loss = \begin{cases}
            S\times L_{seg} + L_{r} + 0.5\times L_{c} & \text{if } S \geq 0.8, \\
            L_{r} + 0.5\times L_{c}   & \text{if } S < 0.8.
            \end{cases}
\end{gather}

\begin{figure*}[!htp]
\centering
\begin{minipage}[b]{0.1\textwidth}
  \centering
  \centerline{\includegraphics[width=2cm]{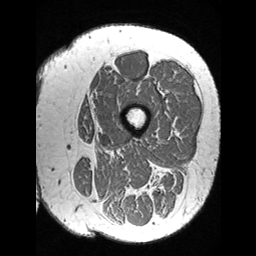}}
  \centerline{(a) Image $X$}\medskip
\end{minipage}
\hfill
\begin{minipage}[b]{0.1\textwidth}
  \centering
  \centerline{\includegraphics[width=2cm]{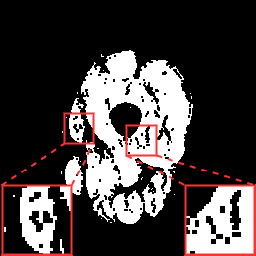}}
  \centerline{(b) GT}\medskip
\end{minipage}
\hfill
\begin{minipage}[b]{0.1\textwidth}
  \centering
  \centerline{\includegraphics[width=2cm]{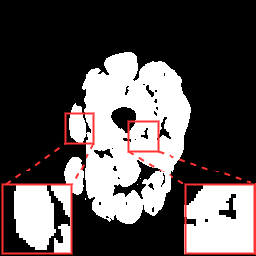}}
  \centerline{(c) Baseline}\medskip
\end{minipage}
\hfill
\begin{minipage}[b]{0.1\textwidth}
  \centering
  \centerline{\includegraphics[width=2cm]{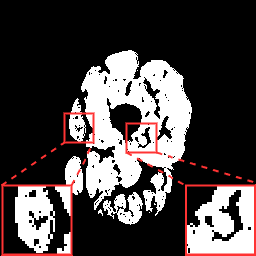}}
  \centerline{(d) \textbf{Our method}}\medskip
\end{minipage}
\hfill
\begin{minipage}[b]{0.1\textwidth}
  \centering
  \centerline{\includegraphics[width=2cm]{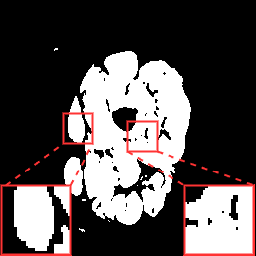}}
  \centerline{(e) ResU-Net}\medskip
\end{minipage}
\hfill
\begin{minipage}[b]{0.1\textwidth}
  \centering
  \centerline{\includegraphics[width=2cm]{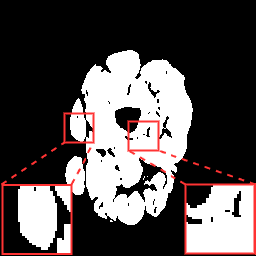}}
  \centerline{(f) AttU-Net}\medskip
\end{minipage}
\hfill
\begin{minipage}[b]{0.1\textwidth}
  \centering
  \centerline{\includegraphics[width=2cm]{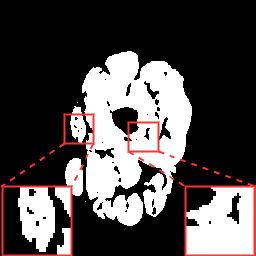}}
  \centerline{(g) MC-Net}\medskip
\end{minipage}
\hfill
\begin{minipage}[b]{0.1\textwidth}
  \centering
  \centerline{\includegraphics[width=2cm]{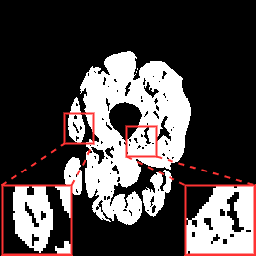}}
  \centerline{(h) Fully U}\medskip
\end{minipage}
\caption{Muscle segmentation results with (a) as the input of the framework. (b) is the ground truth image GT and (c)-(h) are the corresponding muscle masks generated by baseline, the proposed method, ResU-Net, AttentionU-Net (AttU-Net), MC-Net, and fully-supervised U-Net (Fully U), respectively. The interested areas are enclosed with red squares.}
\label{fig:3}
\end{figure*}

\subsection{Data Augmentation}
\label{sec:DA}
While the previous contributions focus on enhancing the dataset with pseudo-labels, we further address the limited amount of data~\cite{chen2020simple} by using augmentation techniques (including contrast adjustments to deal with inhomogeneities) that can help to learn better feature representations for muscles and IMF. Specifically, given an input image $X$, we apply an augmentation function $T$ that consists of flipping, rotation, and affine transformations. Finally, we adjust the contrast of $T(X)$ by a random factor $\gamma\in[0.5, 0.7]$.

\section{Experiments}
\label{sec:experiments}

\subsection{Dataset and Implementation Details}
We evaluate the proposed framework on a thigh dataset from multiple sclerosis subjects. The dataset consists of magnetic resonance imaging (MRI) volumes of the thigh muscle from 11 participants that were scanned at different timelines and precisely annotated (excluding IMF). All subjects were scanned with a 3DT1 sequence (IRFSPGR, TE $= 2.7msec$, TR $= 6.5 msec$, acquisition matrix $= 480 \times 480$, Slice thickness $= 1mm$). The baseline and follow-up images were registered following the process described in Tang et al.~\cite{tang2018automatic} and using the femur as the target. All the volumes were pre-processed into 2D slices of $256\times256$~\cite{muscle_AttnU}. A random subject-wise data split of $7$ and $4$ participants for training and testing, respectively, was applied to the multiple timepoints. The final dataset contained $1040$ slices from $34$ 3D MRIs for training and $249$ slices from $8$ 3D MRIs for testing. 13 of the training slices, around $1\%$ in the training set, were randomly selected as precise annotations for the PLG stage only. Finally, the two stages of our framework were trained for $20$ (PLG) and $10$ (PLD) epochs, respectively. The optimization settings followed Zhu et al.'s training scheme~\cite{zhu2017unpaired}. The code was implemented in PyTorch 1.4 and tested with an NVIDIA GTX 1080.

\subsection{Evaluations and Comparisons}
For the evaluation of the segmentation, we use the Dice similarity coefficient for both thigh muscle ($Dice_{TM}$) and IMF ($Dice_{IMF}$). To generate an IMF mask, we apply a binary closing operation to the testing thigh muscle mask ($Y_{test}$) to close all IMF branches and generate noisy masks ($Y_{noisy}$). The differences between the $Y_{test}$ and $Y_{noisy}$ mask are then considered the IMF mask. The results on that mask can give us a clearer idea on the improvement to exclude IMF areas from the final mask.

Table~\ref{table: 1} lists the quantitative results (evaluated on the testing data) including state-of-the-art approaches for the following methods: U-Net (Baseline)~\cite{Unet}, ResU-Net~\cite{resU}, AttentionU-Net~\cite{AttU,muscle_AttnU}, MC-Net~\cite{McNet}, a fully supervised U-Net, and our proposed method. Qualitative results are shown in Fig.~\ref{fig:3}.  When only $13$ labeled 2D samples were available for training, the proposed method outperformed all the compared methods with an improvement of the $Dice_{IMF}$ and $Dice_{TM}$ metrics of $11\%$ and $4.2\%$ when compared to the baseline. Furthermore, our method proposal with a low number of samples gave results comparable to those of a fully supervised U-Net trained with $1040$ labeled samples.
 \begin{table}[htbp]\centering
        \begin{tabular}{c|cc}
         \hline
        Method & $Dice_{IMF}$ & $Dice_{TM}$\\
        \hline
        Fully supervised U-Net  & \textit{0.772} & \textit{0.940}\\
        U-Net (Baseline)    & 0.573 & 0.864\\   
        ResU-Net & 0.577 & 0.862\\
        AttentionU-Net & 0.574 & 0.862\\
        MC-Net & 0.582 &  0.856\\
        \hline
         \multicolumn{3}{c}{Augmentation} \\
        \hline
        w/o Aug + w/o CA & 0.665 & 0.904 \\
        Aug + w/o CA &  0.671 &  \textbf{0.907} \\
        w/o Aug + CA & 0.673 & 0.904 \\
        \hline
        \multicolumn{3}{c}{PLD stage} \\
        \hline
        w/o CE + w/o $L_{r}$   & 0.654 & 0.879\\
        CE + w/o $L_{r}$  & 0.674 & 0.894\\\hline   
        Our method & \textbf{0.683}  & 0.906\\   
        \hline
        \end{tabular}
        \caption{IMF and muscle segmentation performance summary evaluated on the testing data. The table includes state-of-the-art approaches, an ablation study regarding augmentation and contrast adjustment (CA), and an ablation study (where w/o represents for without) regarding different training strategies for the MC-Net in the PLD stage.}
        \label{table: 1}
 \end{table}

\subsection{Ablation Study}
\begin{figure}[htb]
\centering
\begin{minipage}[b]{0.23\linewidth}
  \centering
  \centerline{\includegraphics[width=2cm]{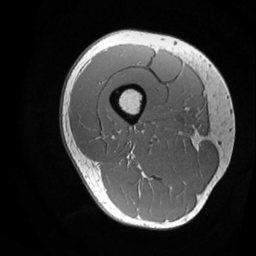}}
  \centerline{(a) Image $X$}\medskip
\end{minipage}
\hfill
\begin{minipage}[b]{0.23\linewidth}
  \centering
  \centerline{\includegraphics[width=2cm]{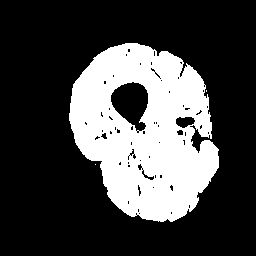}}
  \centerline{(b) GT}\medskip
\end{minipage}
\hfill
\begin{minipage}[b]{0.23\linewidth}
  \centering
  \centerline{\includegraphics[width=2cm]{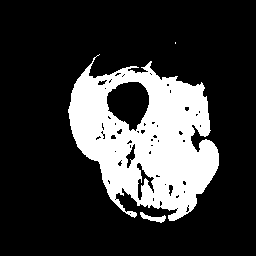}}
  \centerline{(c) Without $L_{r}$}\medskip
\end{minipage}
\hfill
\begin{minipage}[b]{0.23\linewidth}
  \centering
  \centerline{\includegraphics[width=2cm]{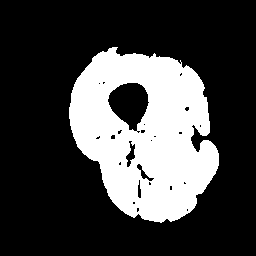}}
  \centerline{(d) With $L_{r}$}\medskip
\end{minipage}
\caption{Qualitative comparison on the effect of $L_r$ on the segmentation results: (a) input image $X$, (b) ground truth image GT, (c) prediction without $L_r$, and (d) prediction with $L_{r}$.}
\label{fig:4}
\end{figure}

To verify the effectiveness of each component in our proposed framework, we conducted a set of ablation studies. Table~\ref{table: 1} summarizes the results regarding the data augmentation strategies. Adjusting the contrast of input images reduces the effect from intensity inhomogeneity as evidenced by the improvement on the Dice metrics. Further ablation studies were conducted regarding the PLD stage. When training the model without confidence evaluation (CE), the performance decreases considerably as $L_{seg}$ is supervised by $Y_{corrected}$ masks that contain potentially incomplete structures. By constraining the supervision signal with CE and $L_{r}$ (visual examples in Fig.~\ref{fig:4}), the model is capable of better exploiting the $Y_{corrected}$ masks.

\section{Conclusion}
\label{sec:conclusion}
In this paper, we propose a two-stage few-shot segmentation framework to generate precise thigh muscle masks without IMF. Combining pseudo-label evaluation and denoising schemes, our method could be trained with only $13$ images ($1\%$ of the dataset) to achieve a comparable performance to a fully supervised method and outperform other state-of-the-art approaches with limited annotations. Therefore, our method presents a novel way to leverage class prototypes in scenarios with scarce and noisy annotations. Moreover, our proposed robust noise loss provides a new supervision approach for exploiting uncertainty in semi-supervised problems. In conclusion, our proposed method can generate precise thigh muscle masks where the IMF is excluded and could be used as part of an automatic muscle volume tracking tool for longitudinal clinical studies.

\section{Compliance with ethical standards}
The study was approved by the University of Sydney Human Research and Ethics Committee and all procedures adhered the tenets of the Declaration of Helsinki.

\section{ACKNOWLEDGEMENTS}
The authors would like to acknowledge funding support of Multiple Sclerosis Research Australia (18-0416),  Australia Medical Research Future Fund under Grant (MRFFAI000085) and Australian Government Research Training Program (RTP) Scholarship.

\small{
\bibliographystyle{IEEEbib.bst}

\end{document}